\documentclass[aps,pra,showpacs,twocolumn,superscriptaddress]{revtex4-1}
\usepackage{graphicx}
\usepackage[usenames]{color}
\usepackage{amssymb,amsmath}

\newcommand{\eq}[1]{Eq.~(\ref{#1})}
\newcommand{\fig}[1]{Fig.~\ref{#1}}
\newcommand{\be}[1]{\begin{equation}\label{#1}}
\newcommand{\ee}{\end{equation}}

\begin{document}

\title{ Frustrated double ionization in two-electron triatomic molecules}

\author{A. Chen}
\affiliation{Department of Physics and Astronomy, University College London, Gower Street, London WC1E 6BT, United Kingdom}
\author{H. Price}
\affiliation{Department of Physics and Astronomy, University College London, Gower Street, London WC1E 6BT, United Kingdom}
\author{A. Staudte}
\affiliation{Joint Laboratory for Attosecond Science, University of Ottawa and National Research Council,
100 Sussex Drive, Ottawa, Ontario, Canada K1A 0R6}
\author{A. Emmanouilidou}
\affiliation{Department of Physics and Astronomy, University College London, Gower Street, London WC1E 6BT, United Kingdom}

\begin{abstract}
Using a semi-classical model, we investigate frustrated double  ionization (FDI) in $\mathrm{D_3^+}$, a two-electron triatomic molecule,  when driven by an intense, linearly polarized, near-infrared (800 nm) laser field. We compute the kinetic energy release of the nuclei and find  a good agreement between experiment and our model.  We explore the two pathways of FDI and show that, with increasing field strength, over-the-barrier ionization overtakes tunnel ionization as the underlying mechanism of FDI. Moreover, we compute the angular distribution  of the ion fragments for FDI and identify a feature that can potentially  be observed experimentally and  is a signature of only one of the two pathways of FDI.   \end{abstract}
\date{\today}

\maketitle

The nonlinear response of multi-center molecules to intense laser fields  is a fundamental problem, posing challenges to  theory and experiment alike.
The starting point of these interactions is often tunnel-ionization in the laser field. Due to the oscillating electric field the ionized electron does not always escape, and can be recaptured by the parent ion into a Rydberg state at the end of the laser pulse. This mechanism is called frustrated ionization \cite{Eichmann1,Eichmann2,nubbe,Doerner2011,Emmanouilidou2012,Emmanouilidou2014,JcKenna}. Frustrated ionization was initially studied in He \cite{Eichmann2} and H$_2$ \cite{Eichmann1}, and is also a candidate for the inversion of neutral N$_2$ in the context of free-space air lasing \cite{lasing}. So far, D$_3^+$ and H$_3^+$ are the only multi-center molecules, where frustrated ionization has been studied in a benchmark experiment  \cite{JMcKenna, JcKenna,JcKenna2} and discussed using classical models \cite{Classic1,Classic2}.

Here, we explore double ionization (DI) and frustrated double  ionization  (FDI) of strongly-driven D$_3^+$. In FDI both electrons ionize but one is recaptured. We model these mechanisms using a three-dimensional (3D) semi-classical trajectory simulation. Our model accounts for tunnel-ionization during propagation thus providing an improved description of FDI---a significant process that accounts for roughly 10\% of all events in triatomic molecules \cite{Classic1,Classic2}. 

Classical and semi-classical models are essential in understanding the fragmentation dynamics in  triatomic molecules driven by intense infrared laser pulses. One reason is 
that the strongly-driven dynamics of two electrons and three nuclei poses an immense challenge for fully ab-initio quantum mechanical calculations. Currently, quantum mechanical techniques can only address one electron in triatomic molecules in two-dimensions \cite{Bandrauk}. Our work employs a 3D semi-classical model which has provided significant insights into the FDI process for strongly-driven $\mathrm{H_{2}}$  \cite{Emmanouilidou2012, Emmanouilidou2014}.  Here, we generalize our model to triatomic molecules. We  show that our result for the distribution of the kinetic energy release (KER) for FDI is in good agreement with the experimental result in ref.\cite{JMcKenna}. Moreover, even though FDI is generally associated with tunnel-ionization, we show that for increasing field strengths the mechanism underlying FDI is over-the-barrier ionization instead.  We have previously shown that two pathways contribute to FDI \cite{Emmanouilidou2012}. We show that for the strongly-driven triatomic  D$_3^+$   one of the two pathways  of FDI has  a signature in the angular distribution of the ion fragments with respect to the direction of the laser field. Very importantly, this trace can potentially be observed experimentally.

In our model we employ a laser field of the form   
\begin{eqnarray}
\begin{split}
&\mathrm{{\bf{E}}(t)=E_{0}(t)\cos({\omega t})\hat{z}}\\
&\mathrm{E_{0}(t)}=\left\{ 
\label{eqnfield}
\begin{array}{lcc}
\mathrm{E_{0}} && \mathrm{0 \leq t < 10T}\\
\mathrm{E_{0}\cos^2\frac{\omega(t-10T )}{8}} && \mathrm{10T \leq t \leq 12T},
\end{array}
\right.
\end{split}
\end{eqnarray}

\noindent  with $\mathrm {E_{0}(t)}$, $\mathrm{T}$ and $\mathrm{\omega}$ the envelope, the  period and the frequency, respectively, of the laser field. We take $\mathrm{\omega}$
 equal to  0.057 a.u.  (800 nm). In the following, we consider only two cases of planar alignment, i.e., one side of the equilateral, molecular triangle is either parallel or perpendicular to the $\mathrm{\hat{z}}$-axis.     Atomic units are used throughout this work unless otherwise indicated.

 To compare with the experimental results  \cite{JcKenna,JMcKenna} we take the initial state of the $\mathrm{D_{3}^{+}}$ molecule  to be the one created via the reaction $\mathrm{D_{2}+D_{2}^{+}\rightarrow D_{3}^{+}+D}$ \cite{JcKenna,JMcKenna}.
  This initial state consists of a superposition of vibrational states ${\it v}=1-12$   \cite{D3_vibrational, JMcKenna}, each with a triangular configuration. 
 Tunnel ionization is very sensitive to variations of the ionization potential and known to preferentially ionize larger internuclear separations \cite{Ergler2006PRL,Goll2006PRL}. Thus, we assume that most of the D$_3^+$ ionization occurs at the outer classical turning point of the vibrational levels. The turning point varies from 2.04~a.u. ($v=1$) to 2.92~a.u. ($v=12$)  \cite{D3_vibrational, D3_potential}.
  We find the first and second ionization potentials of the relevant 12 vibrational states   using the quantum chemistry software MOLPRO \cite{MOLPRO}. For the initial state of $\mathrm{D_{3}^{+}}$ in the laser field, we assume that one electron (electron 1) escapes either by tunneling or over-the-barrier ionization in the field-lowered Coulomb potential \cite{Emmanouilidou2014}, depending on the field strength and the vibrational state.   We use a tunneling rate given by the semi-classical formula in ref.\cite{Murray2011}. If electron 1 escapes  by tunneling then  its transverse to the laser field velocity  distribution  is a Gaussian \cite{ADK}, while its velocity  parallel to the laser field is assumed to be zero. This assumption has been recently verified experimentally for strongly-driven Ar  \cite{Moshammer}.  We assume that the other electron is initially bound (electron 2).  Its initial state is described by a 
 microcanonical distribution  which we formulated very recently  \cite{microcanonical}  and is applicable to any one-electron triatomic molecule.  Since an initial pre-dissociation does not significantly modify the ionization dynamics \cite{Emmanouilidou2014}, we simplify our model  by initializing the nuclei at rest for all vibrational levels. 
 
 For the propagation of our model system we generalized the technique we developed in the context of FDI of H$_2$. Details can be found in ref.\cite{Emmanouilidou2014}.
    It is worth emphasizing a feature of our model:  we use the Wentzel-Kramers-Brillouin approximation  to allow for tunneling of each electron during the propagation  \cite{Emmanouilidou2012, Emmanouilidou2014}.  This is essential in order for our model to accurately describe the enhanced ionization process  (EI) \cite{Niikura,bandrauk1996}. In EI,  at a critical distance of the nuclei, a double potential well is formed such that it is easier for an electron bound to the higher potential well to tunnel to the lower potential well and subsequently ionize. We also note that our model fully accounts for the Coulomb singularity \cite{Emmanouilidou2014}.

 We now consider  DI and FDI of $\mathrm{D_{3}^{+}}$. DI refers to the formation of three $\mathrm{D^{+}}$ ions and two escaping electrons. FDI   refers to the  formation of a neutral excited fragment $\mathrm{D^{*}}$, two $\mathrm{D^{+}}$ ions and one escaping electron.  Previous experiments on strong-field ionization of  $\mathrm{D_{3}^{+}}$ studied, among other observables, the kinetic energy release (KER), i.e., the sum of the kinetic energies of the ion fragments  \cite{JMcKenna}. To be able to compare the experimental   KER with the KER from our simulation  we need to account for the  intensity averaging in the focal volume.  
 
 We first compute the KER distribution  for a process $\mathrm{\alpha}=$FDI,DI  as a function of the laser field intensity as follows
 \begin{eqnarray}
 &\mathrm{P^{\alpha}(I,KER)=\frac{\sum\limits_{{\it v},\phi_{0}}{P_{\it v} P^{\alpha}(\phi_{0},{\it v},I,KER)\Gamma(\phi_{0},{\it v},I)}}{\sum\limits_{{\it v},\phi_{0}}{P_{\it v} \Gamma(\phi_{0},{\it v}, I)}}}&,
 \end{eqnarray}
where $\mathrm{P^{\alpha}(\phi_{0},{\it v},I,KER)}$ is the probability  to obtain a KER from a  vibrational state {\it v}, for an  initial phase of the laser field $ \mathrm{\phi_{0}=\omega t_{0}}$, and for a laser field intensity I.  $\mathrm{I=1/2c \epsilon_{0}E_{0}^2}$, where c is the speed of light and $\mathrm{\epsilon_{0}}$ is the vacuum permittivity. The initial phase $\mathrm{\phi_{0}}$ corresponds to the starting point of the propagation.      $\mathrm{\Gamma(\phi_{0},{\it v},I)}$ is the  tunnel-ionization rate computed using the semi-classical formula in ref.\cite{Murray2011} and $\mathrm{P_{\it v}}$ is the percentage of the vibrational state {\it v} in the initial state produced following the reaction generating $\mathrm{D_{3}^{+}}$ \cite{D3_vibrational}.
\begin{figure}[h]
  \begin{center}
  \centering
 \includegraphics[clip,height=0.35\textwidth]{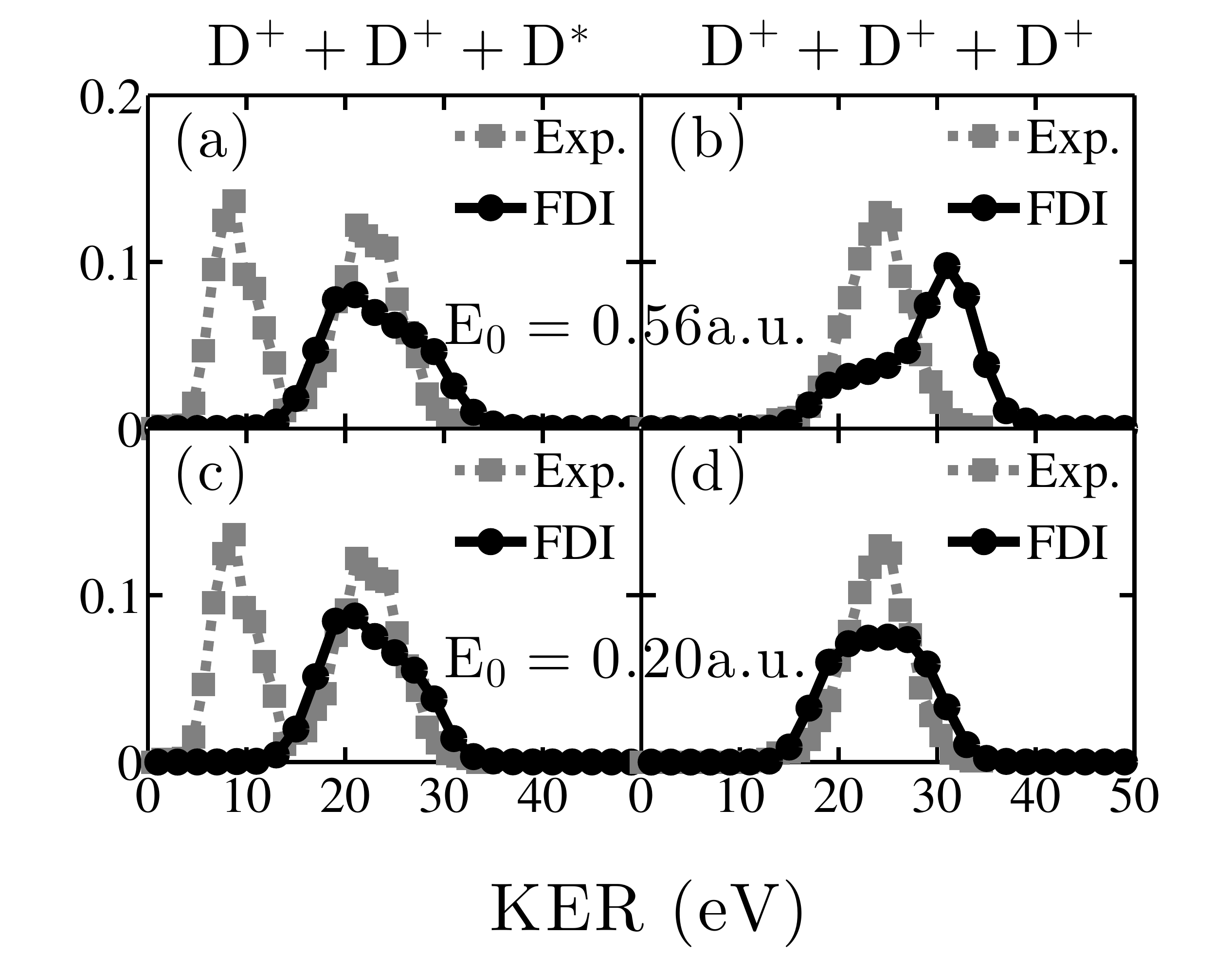}
  \end{center}
\caption{  Intensity averaged KER distributions for FDI  and DI for $\mathrm{I_{max}}$ corresponding to a field strength of $\mathrm{E_{0}=}$0.56 a.u., (a) and (b), and of  $\mathrm{E_{0}=}$0.2 a.u., (c) and (d).   The grey dashed lines show the relevant experimental results from \cite{JMcKenna}. Our results and the experimental ones for DI and FDI   have been normalized to 1. Note that the experimental results in (a) and (c) have two peaks; it is the area under the  higher energy peak that has been normalized to 1.  } 
\label{KER}
\end{figure}
 Following the formulation  in \cite{laser_volume_effect} we  compute the KER distribution  for a laser peak intensity I$_{max}$ as follows

 \begin{eqnarray}
\mathrm{P^{\alpha}(I_{max}, KER)=\int_0^{I_{max}} \frac{P^{\alpha}(I,KER)}{I}dI}.
\label{average}
\end{eqnarray}

In practice,  in \eq{average}  we  integrate only over the intensities which significantly contribute to the process $\mathrm{\alpha}$. To find the lower limit of these intensities we compute the ionization probability for an intensity I and a vibrational state {\it v}, which for small values of the ionization probability is given by

\begin{eqnarray}
\mathrm{\Gamma({\it v}, I)}\approx\int^{t_f}_{t_i}\mathrm{\Gamma(\omega t, {\it v}, I)}dt.
\end{eqnarray}
with the integration over the duration of the laser pulse.   For the laser pulse and all the vibrational states  of the triatomic we currently consider, we find that the ionization probability  of $\mathrm{D_{3}^{+}}$ is very small for  field strengths less than 0.06 a.u..  Therefore, only field strengths above 0.06 a.u. contribute to the observed KER distributions. 

We now compare the intensity-averaged KER distributions with the measured ones \cite{JMcKenna}  for a peak field strength of 0.56 a.u., which  corresponds to the experiment's intensity of 1.1$\times$10$^{16}$ W/cm$^2$  \cite{JMcKenna}. 
 We find that the KER distributions for FDI and DI for both considered alignments of the molecule relative to the laser field polarization are very similar. We therefore expect that any other planar alignment of molecule and laser field polarization  will not significantly change the KER distributions. We plot the KER  in \fig{KER} only  for the parallel alignment.
We find that for FDI the computed KER distribution  is in good agreement with the experimental one, see \fig{KER} (a). Both distributions peak at 21 eV, while the computed KER distribution  has a wider tail towards higher field strengths.
In the experimental data of the single ionization channel (Fig. 1 (a) and (c)) an additional peak at $\approx$8~eV is present. This peak is likely due to the bondsoftening \cite{Bucksbaum1990PRL} of an intermediate D$_2^+$ in the experiment. Our model does not include this mechanism and hence, does not show this peak.
 
\begin{figure} [ht]
\centering
 \includegraphics[clip,height=0.17\textwidth]{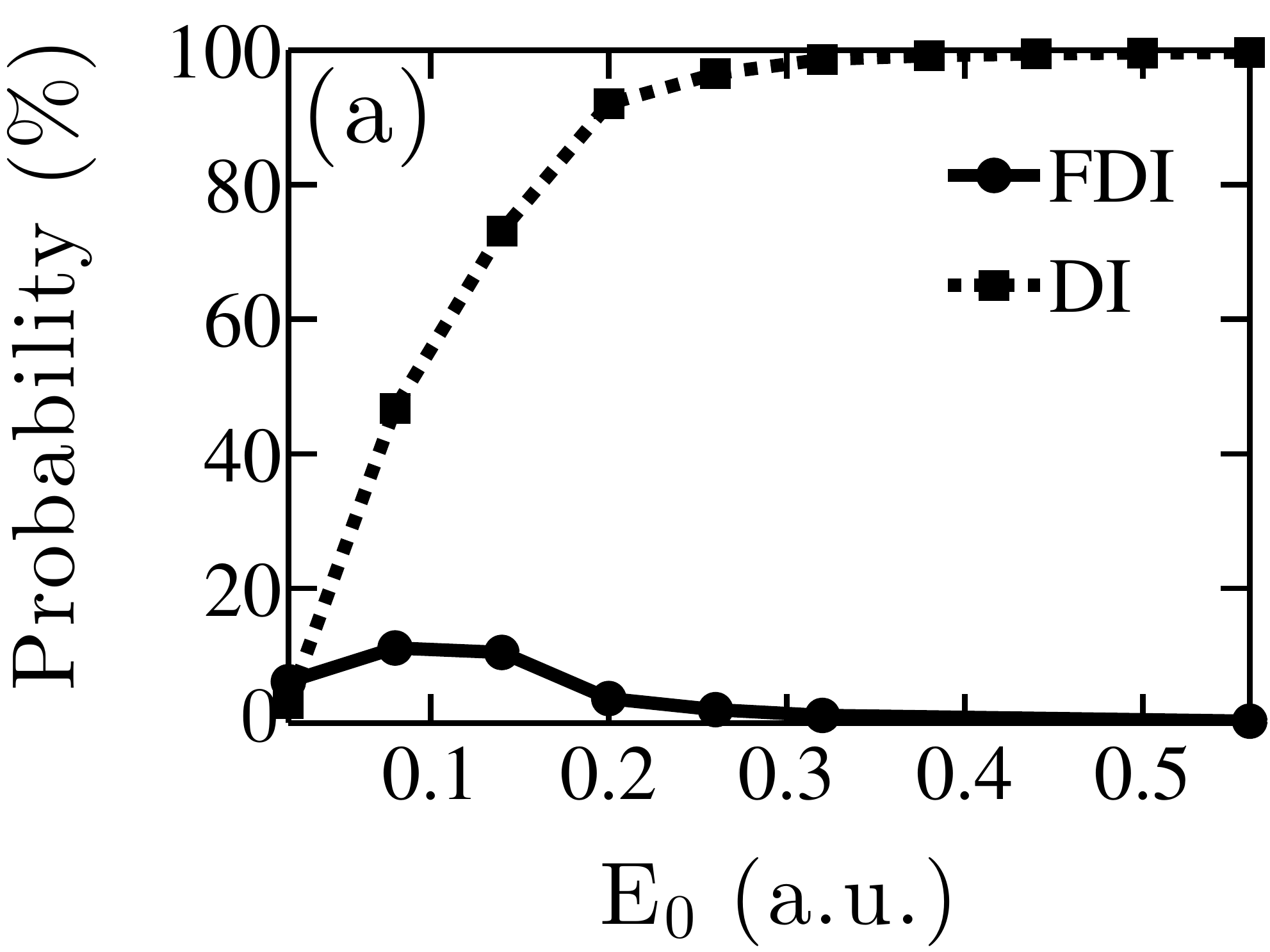}
 \includegraphics[clip,height=0.17\textwidth]{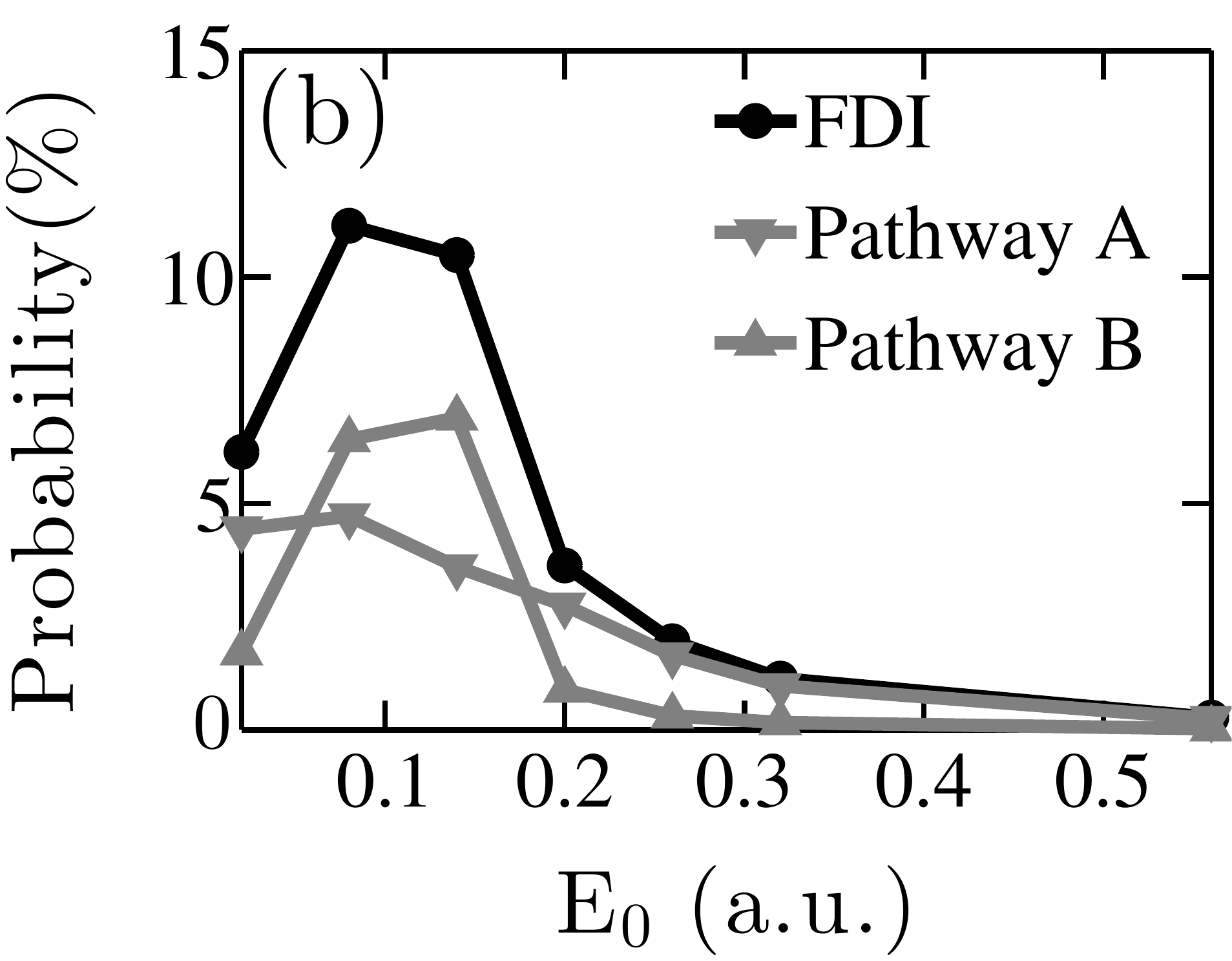}
\caption{ The  probability  (a) for FDI and DI and (b) for pathways A and B of FDI as a function of the field strength $\mathrm{E_0}$. The smallest strength of the laser field we consider in this figure is $\mathrm{E_0}=0.02$ a.u.. }
\label{probFDI&DI}
\end{figure}

The agreement is not as good for the KER distribution  for DI shown in \fig{KER} (b): the computed  distribution peaks at 31 eV while the experimental one peaks at 24 eV. It is possible  that
our model overestimates the DI probability for high field strengths, see discussion for \fig{probFDI&DI}. Indeed,   when we  consider a smaller peak field strength  of 0.2 a.u. we find that the intensity averaged KER distributions  for FDI and  DI, 
shown in \fig{KER} (c) and (d), respectively, are both in better agreement with the experimental ones at higher intensity.  We note, that our results  compare better with experiment than  previously obtained classical  results \cite{Classic1, Classic2}. In those previous simulations   the KER distributions for DI peak at considerably higher  energies.

To find the upper limit of intensities that contribute to the KER distributions in \fig{KER},  we analyze in \fig{probFDI&DI} the FDI and DI probabilities as a function of the laser field strength. 
In this context, probability is the number of FDI or DI events relative to the number of initialized trajectories. At each intensity we ran enough trajectories to obtain at least 16,000 FDI events and at least 50,000 DI events.
Therefore, the statistical error of these results  is very small. 

\fig{probFDI&DI} (a) shows that the DI probability increases quickly as a function of the field strength reaching already a probability of 99.2\% at a field strength of 0.38 a.u.. Thus,  all  field strengths up to the peak intensity of 0.56 a.u. contribute significantly to the intensity averaged KER distribution for DI in \fig{KER} (b). On the other hand, \fig{probFDI&DI} shows that the FDI probability reaches a maximum of 11.1\% at an intermediate field strength of 0.08 a.u. and then decreases to  0.3\% at a field strength of   0.56 a.u.. Combined with the  1/I factor in \eq{average}, we find that only field strengths up to roughly 0.32 a.u. contribute significantly to the intensity-averaged KER distribution  for FDI in \fig{KER} (b). 

  We now focus on  describing in detail the FDI process for $\mathrm{D_{3}^{+}}$.   
  Similar to the case of H$_{2}$ in  \cite{Emmanouilidou2012} we identify two pathways that can lead to FDI. In the following we refer to the initially tunnel-ionized electron as electron 1 and to the initially bound electron as electron 2. In pathway A, electron 1  escapes, while electron 2 tunnel-ionizes later  while the laser field is on and is eventually  recaptured to a highly-excited state of a D atom. In pathway B, 
electron 1  is eventually recaptured to a highly excited state of D, while electron 2 tunnel-ionizes later but eventually escapes.  We  find  that the distribution of the inter-nuclear distances at the time electron 2 tunnel-ionizes  peaks around 3 a.u. for  $\mathrm{D_{3}^{+}}$. It is mainly after  electron 2 tunnel-ionizes that the nuclei rapidly dissociate, since tunnel-ionization of electron 2  reduces  the screening of the nuclei.   It follows that the KER distribution for FDI should peak at 3 (number of nuclei)/3 (most probable nuclear distance for Coulomb explosion) a.u. 
which is 27.2 eV. Indeed,we find the peak of the computed KER distribution   for FDI to be around 21 eV.  This value is  smaller than 27 eV as expected since one electron in FDI events remains bound screening the Coulomb repulsion of the nuclei.

 In  \fig{probFDI&DI} (b) we show that the probability of pathway B of FDI  reaches a maximum  of 6.9\% at a field strength  of 0.14 a.u. and then decreases fast, reaching less than 1\% at a field strength of 0.2 a.u.. The dominance of pathway B at intermediate intensities is due to electron-electron correlation being much more prominent  for these intensities \cite{Emmanouilidou2009}. Electron-electron correlation was shown to be more important for  pathway B compared to pathway A also for strongly-driven H$_{2}$  \cite{Emmanouilidou2012}. This is to be expected since in pathway B electron 1, following tunnel-ionization, later returns to the ion and interacts with electron 2.   In addition, we  find that, for $\mathrm{D_{3}^{+}}$,  at high intensities, the probability for pathway A of FDI  decreases at a much slower rate than the probability of pathway B, see \fig{probFDI&DI} (b). The reason is that 
 in pathway A  electron 2 tunnels after gaining energy in a frustrated enhanced ionization process, i.e.,   electron 2 gains energy from the field in the same way as in an  enhanced ionization process  \cite{Niikura,bandrauk1996} but electron 2 eventually  does not escape.
 For higher intensities electron correlation plays a less important role compared to enhanced ionization. Hence, the probability for pathway A reduces at a smaller rate than the probability for  B.

  Next, we will identify the prevalent ionization mechanism leading to FDI in D$_3^+$. Specifically,  we determine the probability of over-the-barrier ionization, $\mathrm{P_{OBI}}$.
In our notation $\mathrm{P_{OBI}}$ not only refers to the permanent ionization of electron 2 in pathway B but also includes the temporary ionization of electron 2 in pathway A before it is being recaptured into an excited D$^*$ state.
We find that $\mathrm{P_{OBI}}$ is around 9\% for a field strength of 0.08 a.u.  increasing to  87\% at   0.56 a.u..  
 However, for field strengths above 0.32 a.u., the probability of FDI events reduces significantly, see \fig{probFDI&DI} (a). Therefore,  after integrating over field strengths up to 0.56 a.u., using \eq{average}, we find that over-the-barrier ionization accounts for 21\% of FDI events. Thus, tunnel-ionization dominates FDI.


\begin{figure} [ht]
\centering
 \includegraphics[clip,width=0.480\textwidth]{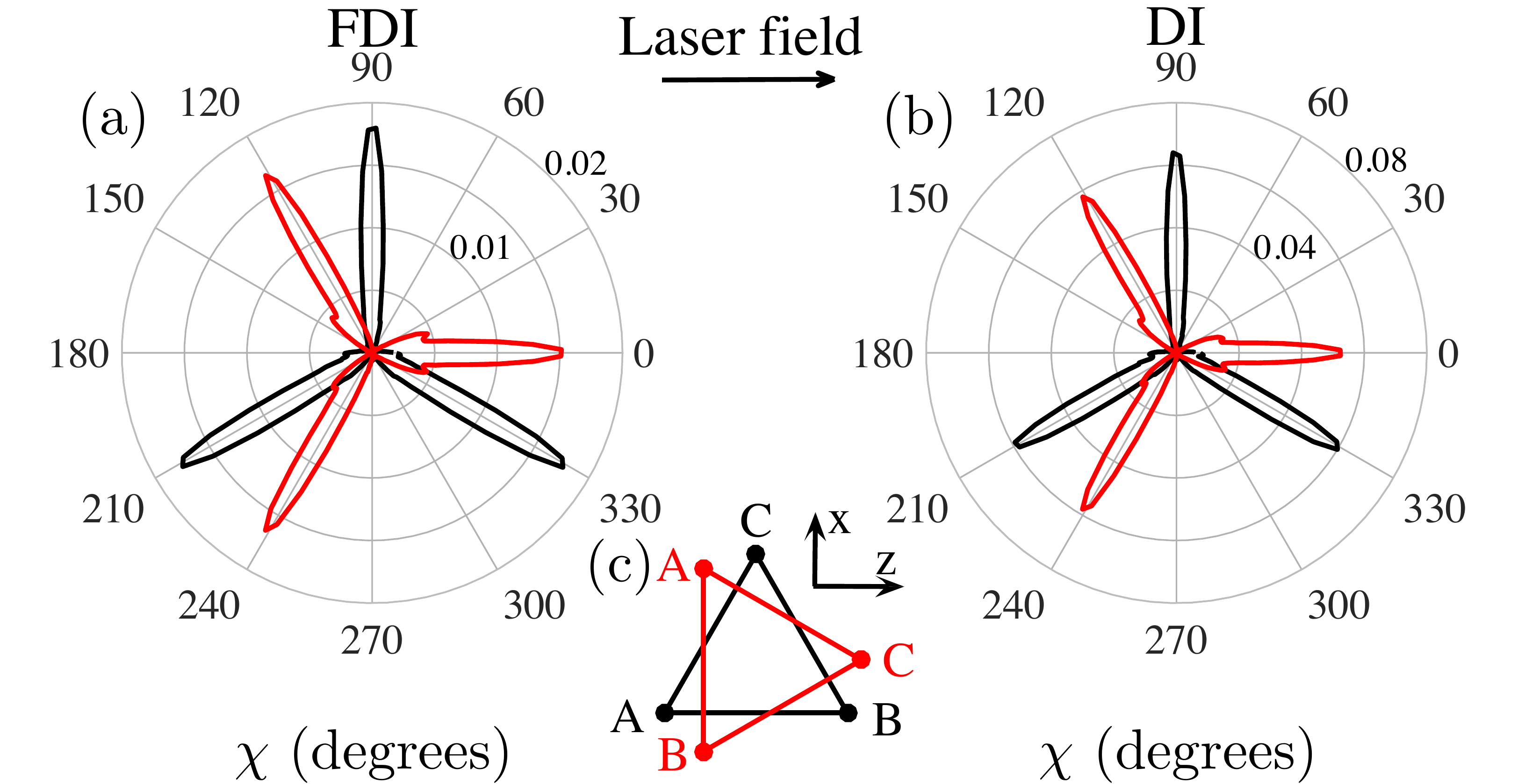}
\caption{(color online)   The angle of the velocity vector of D$^{*}$ in FDI (a) and of D$^{+}$ in DI (b) with respect to the laser field for parallel  (black) and for perpendicular  (red) alignment. The field strength is 0.08 a.u..(c) The initial state geometric configuration of D$^+_3$ with respect to the laser field.}
\label{angles}
\end{figure}

\begin{figure} [ht]
\centering
 \includegraphics[clip,width=0.2300\textwidth]{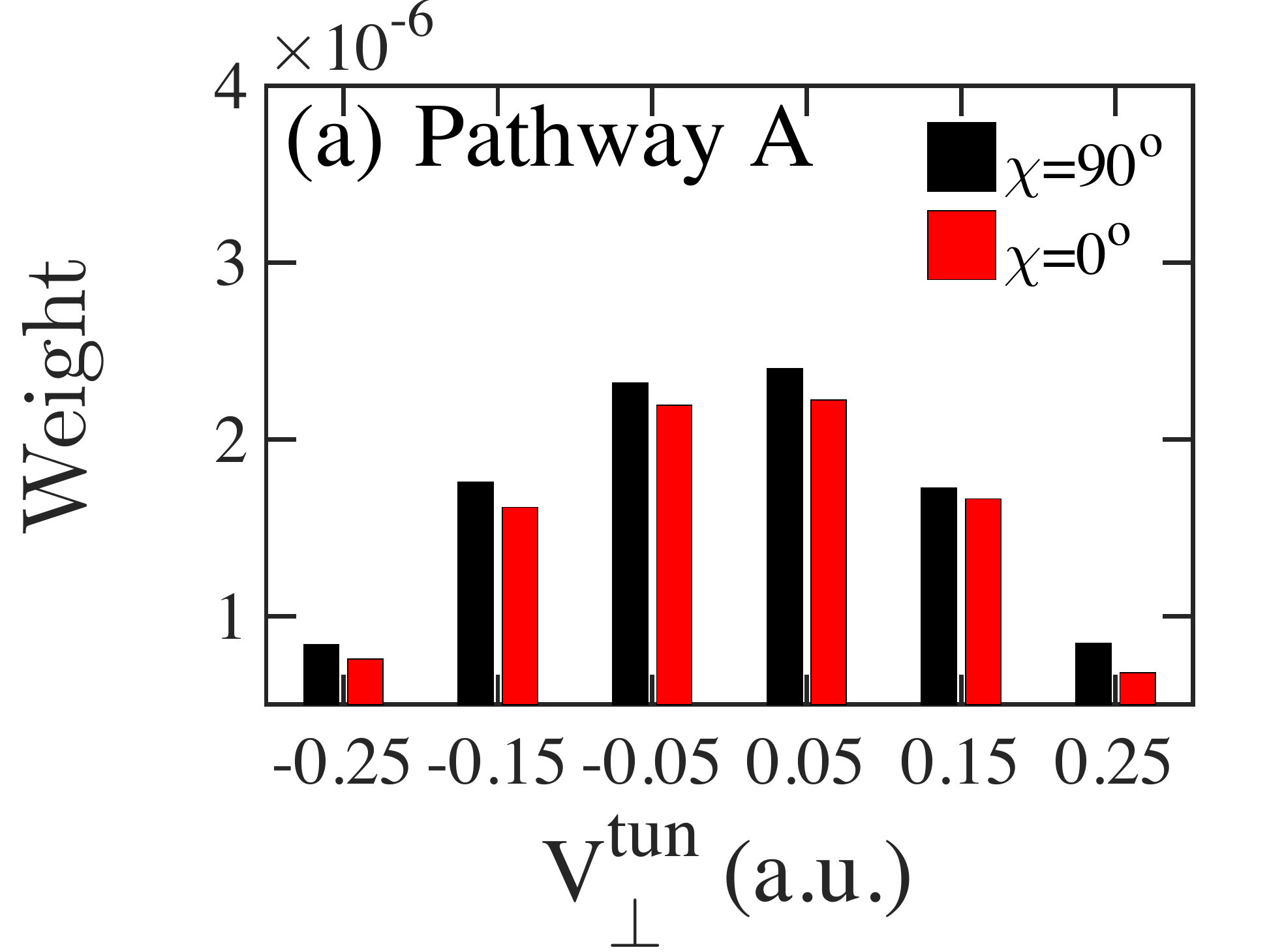}
 \includegraphics[clip,width=0.2300\textwidth]{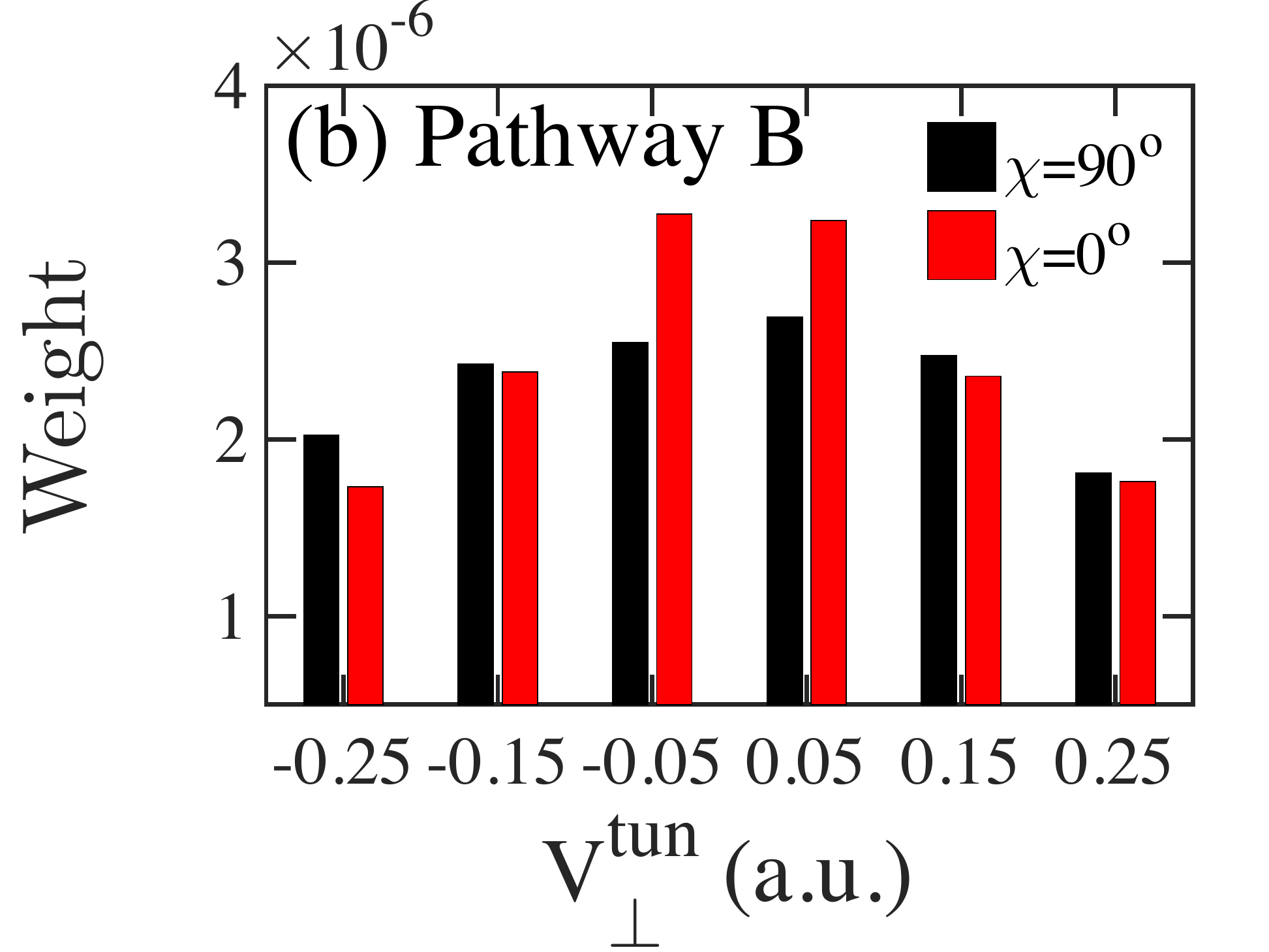}
\caption{(color online) The weight of the lobes around  0$^{\circ}$ and 90$^{\circ}$ for pathway A (a) and B (b). }
\label{angles1}
\end{figure}



Finally, we  identify  a feature of the break-up dynamics of the strongly-driven triatomic that is a signature of pathway B and can be potentially observed experimentally. In \fig{angles} (a)/(b) we plot for FDI/DI  the  angle $\chi$ of the  velocity of D$^{*}$/D$^{+}$ with respect to the  laser field for a field strength of 0.08 a.u. for the two alignments of the molecule with respect to the laser field   considered.  We find that, as for DI,  for FDI the angular distribution has a three-lobe structure. The three-lobe structure we obtain for DI is in agreement with previous experiments \cite{JcKenna}. For FDI we cannot provide a direct comparison with experiment since the analyzed data in \cite{JcKenna} includes the angular distribution of all single ionization events D$^{+}$+D$^{+}$+D, i.e., FDI events as well as bondsoftening events that yield the low KER peak in \fig{KER} (a) and (c). Unlike DI, we find that the three lobes in \fig{angles} (a) do not have equal weight as is the case for DI in \fig{angles} (b). Specifically, the lobe around 0$^{\circ}$  has a 2\% higher weight than the other two lobes  for perpendicular alignment and the lobe around 90$^{\circ}$ has a 2\% less weight than the other two lobes for parallel alignment  in \fig{angles} (a). With respect to the initial state geometry of  the nuclei of $\mathrm{D_{3}^{+}}$  in \fig{angles} (c), this means
that  the electron that finally stays bound in FDI gets attached for parallel alignment more   to either nucleus A or B rather than  C 
while for perpendicular alignment   to nucleus C. This difference is reasonable since frustrated enhanced ionization takes place mainly between the nuclei that are more parallel to the field, A and B for parallel alignment and A and C or B and C for perpendicular alignment.  We find that this small difference in the weight of the lobes is present in both pathways A and B. 

We find that  the probability of the electron that finally remains bound in FDI  to get attached to different nuclei varies significantly  as a function of the initial velocity of the tunneling electron. For pathway A as a function of the initial velocity of electron 1, we find that  the probability for electron 2 to get attached to nucleus C is  between 1.5 \% and 2 \% smaller/larger than the probability to get attached to nuclei A and B for parallel/perpendicular  alignment. Thus, the probability  of electron 2  to get attached to nuclei A, B and C is not sensitive to the initial velocity of electron 1. However, we find that for pathway B the probability for electron 1 to get attached to nuclei A, B and C varies  with the initial velocity of electron 1 in the direction perpendicular to the field. Namely, for parallel alignment, we find that it is for small initial velocities of electron 1 that the probability of electron 1 to get attached to nucleus C 
differs the most from the probability of getting attached to nuclei A or B; the probability to get attached to nucleus C is roughly 7 \% smaller. For perpendicular alignment, we find  again that it is for small initial velocities of electron 1 that the probability of electron 1 to get attached to nucleus C differs the most from the probability of getting attached to nuclei A or B; the probability to get attached to nucleus C is roughly 7 \% larger. Expressing  the above differences in terms of the lobes of the angular distribution of FDI it means that, for pathway B,  for very small initial velocities of electron 1 the lobe around 90$^{\circ}$ has 7 \% less weight than each of the other two lobes of parallel polarization while the lobe around 0$^{\circ}$ has 7 \% more weight than each of the other two lobes of perpendicular polarization. Thus, if the initial velocity of electron 1 can be probed experimentally then one would observe a significant difference in the weight of the lobes around 0$^{\circ}$ and 90$^{\circ}$ that is due to pathway B. We illustrate the latter  in \fig{angles1}. In \fig{angles1} (a), we show  that in pathway A the difference in weight between the lobes around 0$^{\circ}$ and 90$^{\circ}$ is small and insensitive to the initial velocity of electron 1. In contrast, in \fig{angles1} (b), we  show that for pathway B the difference in weight between the lobes around 0$^{\circ}$ and 90$^{\circ}$  is very sensitive to the initial velocity of electron 1 and is large for small initial velocities.

Concluding, using a 3D semi-classical calculation where the Coulomb singularity is fully accounted for, we find that  our results for the KER distribution  for FDI of the  strongly-driven $\mathrm{D_{3}^{+}}$ agree well with experimental results. We also find 
that the underlying mechanism for FDI switches from tunnel to over-the-barrier ionization  with increasing field strength. It would be interesting if future experiments identify the asymmetry in the angular distribution of the D$^{*}$ fragment for FDI events which 
we have shown to be a signature of pathway B of FDI.

{\it Acknowledgments.} A.E acknowledges  I. Ben-Itzhak and  J. McKenna for re-analyzing their published data in order  to enable comparison with our simulations as well as many discussions with I. Ben-Itzhak. A. E also   acknowledges the EPSRC grant no. J0171831 and the use of the computational resources of Legion at UCL.

\end{document}